\documentstyle[prl,aps,multicol,epsf]{revtex}
\renewcommand{\narrowtext}{\begin{multicols}{2}
\global\columnwidth20.5pc} 
\renewcommand{\widetext}{\end{multicols}
\global\columnwidth42.5pc} \multicolsep = 8pt plus 4pt minus 3pt

\begin{document}
\draft

\title{Fast Response and Temporal Coding on Coherent Oscillations
in Small-World Networks}

\author{Luis F. Lago-Fern\'{a}ndez$^{(1)}$,
Ram\'{o}n Huerta$^{(1,2)}$,
Fernando Corbacho$^{(1)}$,
and Juan A. Sig\"{u}enza$^{(1)}$}

\address{(1) Grupo de Neurocomputaci\'{o}n Biol\'{o}gica (GNB),
E.T.S. de Ingenier\'{\i}a Inform\'{a}tica,
Universidad Aut\'{o}noma de Madrid, 28049 Madrid (SPAIN).}

\address{(2) Institute for Nonlinear Science,
University of California, San Diego, La Jolla, CA  92093-0402}

\date{\today}

\maketitle

\begin{abstract}
We have investigated the role that different connectivity regimes 
play on the dynamics of a network
of Hodgkin-Huxley neurons by computer simulations.
The different connectivity topologies exhibit the
following features: 
random connectivity topologies give rise to fast 
system response yet are unable to produce coherent oscillations in 
the average activity of the network;
on the other hand, regular connectivity topologies give rise to
coherent oscillations and temporal coding, but in a temporal scale 
that is not in accordance with fast signal processing.
Finally, small-world (SW) connectivity topologies, which fall
between random and regular ones, 
take advantage of the best features of both,
giving rise to fast system
response with coherent oscillations along with reproducible
temporal coding on clusters of neurons. 
Our work is the first, to the best of our knowledge, to show
the need for a small-world topology in order to obtain
all these features in
synergy within a biologically plausible time scale.
\end{abstract}


\narrowtext

In a recent letter by Watts and Strogatz \cite{a1} it was shown that 
small-world networks enhance signal-propagation speed,
computational power, and synchronizability. Small-world stands for
a network whose connectivity topology is placed somewhere between a
regular and a completely random connectivity. The main properties of 
these specific networks are that they can be highly clustered like 
regular networks and, at the same time, have small path lengths like 
random ones.
Therefore, small-world networks may have properties given
neither in regular nor in random networks \cite{barrat,sw1,sw2,sw3}.

In this letter we have extended Watts and Strogatz's general
framework by introducing dynamical elements in the network nodes.
Our source of inspiration is based on a phenomena observed in the 
olfactory antennal lobe (AL) of the locust
discovered by Gilles Laurent and collaborators \cite{a2,a3,a5,a6}.
The AL is a group of around 800 neurons whose functional role 
is to relay information from the olfactory receptors to
higher areas of the brain for further processing. Three main features
have been observed in the dynamics of the AL. First, there is a fast 
response of the AL when the stimulus is presented.
Second, when an odour 
is presented to the insect, coherent oscillations of 20 Hz
in the local field potential (LFP) are measured \cite{a5}. Third, 
every neuron responds to the odour with some particular timing with 
respect to the LFP \cite{a2}. Summarizing: fast response of coherent 
oscillations along with
temporal coding are observed. There are also other systems in the
brain that present coherent LFP oscillations, hence, hinting
to the generality of these phenomena (see \cite{fer} for a review).

The cooperative behavior of large assemblies of dynamical elements
has been the subject of many
investigations\cite{i1,i2,i4,i6,i7,i10,i11,d1,Fohlmeister}. 
In all of them the
connectivity between the elements of the network
was either regular (local or global all-to-all), or random.
However, none of these studies incorporates a comparative analysis
of network dynamics for all the different connectivity topologies.
 
In the present work we pretend to show that in order to provide
fast response, coherent oscillations and temporal coding a 
small-world topology is required. We will show that the regular 
connectivity topology
provides a slow response to the external input. Although it is able
to produce temporal coding and coherent oscillations, the time 
of formation of the oscillations would imply much slower responses
than those observed in biological systems.
On the other hand, for the completely random
connectivity case the responsiveness of the system is highly increased
and temporal variations in clusters activity are present,
but the coherent oscillations tipically observed in the LFP
are lost. Without these coherent oscillations the AL seems to lose
its ability to process the information incoming from 
the sensors \cite{a3}.

The model we propose for this study is made of an array 
of non-identical Hodgkin-Huxley elements coupled by 
excitatory synapses. The unit dynamics is
described by the following set of coupled
ordinary differential equations:

\begin{equation} 
C_{m}\dot{V}_{i} = I^{e}(t) - g_{L}\hat{V}_{L} -
g_{Na}m^{3}h\hat{V}_{Na} - g_{K}n^{4}\hat{V}_{K} + I^{s}(t)
\end{equation}

\begin{mathletters}
\begin{equation}
\dot{m} = \alpha_{m}(V)(1-m) - \beta_{m}(V)m
\end{equation}

\begin{equation}
\dot{h} = \alpha_{h}(V)(1-h) - \beta_{h}(V)h
\end{equation}

\begin{equation}
\dot{n} = \alpha_{n}(V)(1-n) - \beta_{n}(V)n
\end{equation}
\end{mathletters}

\noindent where $V_{i}$ represents the membrane potential of unit 
$i$; $C_{m}$ is the membrane capacitance per unit area; 
$I^{e}(t)$ is the external current,
which occurs as a pulse of amplitude $I_{0}$;
$I^{s}(t)$ is the synaptic current;
$\hat{V}_{r} = V_{i} - V_{r}$, where $V_{r}$ are the equilibrium
potentials for the different ionic contributions ($r = L,Na,K$),
and $g_{r}$ are the corresponding maximum conductances 
per unit area; $h$, $m$, $n$ are the 
voltage dependent conductances;
and $\alpha$, $\beta$ are functions of $V$ 
adjusted to physiological data by voltage clamp techniques.
We have used the original functions and parameters employed by 
Hodgkin and Huxley \cite{hhux}.

The system was integrated using the Runge-Kutta 6(5) scheme with
variable time step based on \cite{11}. The absolute error was 
$10^{-15}$ and the relative error was $10^{-7}$ in all the 
calculations presented in this letter.
The synaptic current $I^{s}$ is given by

\begin{equation}
I^{s}_{i} (t) = g_{ij} r_{j}(t) [V_{s} (t) - E_{s}]
\end{equation}

\noindent where $i$ stands for the index of the neuron that receives 
the synaptic input, $j$ is the neuron from which the synaptic input 
is received, and $g_{ij}$ is the maximum conductance, which 
determines the degree of coupling 
between the two connected neurons. $V_{s}$ is the
postsynaptic  potential, $E_{s}$ is the
synaptic reversal potential and $r_{j}(t)$ is the fraction of bound
receptors computed following the method and parameters described by 
Destexhe et al. \cite{Destx}. Namely, the dynamics of the bound 
receptors $r$ is given by the equation:

\begin{equation}
\dot{r} = \alpha[T](1 - r) - \beta r 
\end{equation}

\noindent where $[T]$ is the concentration of the transmitter,
and $\alpha$, $\beta$ are the rise and decay constants, respectively.

In this model three different kinds of connectivity patterns
have been tested: regular, random and small world. 
To interpolate between regular and random networks we follow the
procedure described by Watts and Strogatz \cite{a1} which we summarize
here for convenience: we start from a ring lattice with $N$ vertices
and $k$ edges per vertex, and each edge is rewired at random with
probability $p$. The limits of regularity and randomness are for 
$p = 0$ and $p = 1$ respectively, and  the small-world
topology lies somewhere in the intermediate region $0 < p < 1$.
The quantification of the structural properties of these graphs is 
performed, following Watts and Strogatz \cite{a1}, using their 
characteristic path length $L(p)$ and their clustering coefficient 
$C(p)$. $L(p)$ is defined as
the number of edges in the shortest path between two vertices,
averaged over all pairs of vertices. $C(p)$ is defined as follows:
suppose that a vertex $v$ has $k_{v}$
neighbours; then at most $k_{v}(k_{v}-1)/2$ edges can exist between 
them. Let $C_{v}$ denote the fraction of these allowable edges 
that actually exist, and define $C$ as the average of $C_{v}$ over 
all vertices $v$. Fig.\ \ref{fig1}a replicates that of Watts and 
Strogatz \cite{a1} for ease of
reference and to verify our computations.

Next we investigate the functional significance of SW topologies for
the dynamics of the network. Watts and Strogatz \cite{a1} already 
note that small-world networks of coupled phase
oscillators synchronize almost as readily as in the mean-field model,
despite having orders of magnitude fewer edges.
To study the global behavior of the network we compute its average
activity $\overline{V(t)} = (1 / N) \sum_{i=1}^{N}
V_{i}(t)$. The quantity
that we use to detect the onset and degree of coherent oscillations is
the average activity oscillation amplitude \cite{i11}, defined by

\begin{equation}
\sigma^{2} (p) = \frac{1}{T_{2}-T_{1}} \int_{T_{1}}^{T_{2}} 
[\langle \overline{V_{p}(t)}\rangle_{t} - \overline{V_{p}(t)}]^{2} dt
\end{equation}

\noindent where $\overline{V_{p}(t)}$ is the average activity of the 
network for a given value of the probability $p$, and the angle 
brackets denote temporal average over the integration interval.
A high value of $\sigma(p)$ would imply a high amplitude of the 
oscillations of the average activity, while a low value would indicate
an almost non-oscillatory behavior.
In Fig.\ \ref{fig1}b we plot $\sigma(p)$ for each of the different 
networks characterized by its probability  p.
Notice that coherent oscillations increase in the region in which 
a high $C(p)$ and a low $L(p)$ occur simultaneously; this is precisely
the SW region. This can be better observed in Fig.\ \ref{fig2}, which
shows the average
activity of the network in three cases corresponding to the
three different topological configurations: regular, random and SW. 
Both the regular and the SW topologies display coherent oscillations,
but in the regular network they appear much later and their amplitude
is smaller than in the SW case.
On the other hand, the random network only displays irregular 
variations over an almost constant pattern of activity.

A more extensive study in the $(k,p)$ plane
has been performed in order to stablish the limits for the 
appearance of coherent oscillations and to check that our
previous results can be generalized within a certain range of 
parameters. We have computed the average activity oscillation 
amplitude for a total of $180$ points in the $(k,p)$ plane, taking 
an integration interval between $T_{1} = 100$ and $T_{2} = 200$. 
An interpolation of these results is plotted in Fig.\ \ref{fig3}, 
where the clear zones indicate high values of $\sigma$. We can 
conclude from this figure that fast coherent oscillations appear 
only in the region of intermediate probabilities, that is, the SW. 
The a priori limits on $k$ are based on the fact that for $k$ lower 
than $\sim10$ the activation of the network is very weak, while for 
$k$ higher than $\sim35$ some neurons become saturated.  

Having shown the necessity of a SW network to obtain a fast response
along with coherent oscillations of the average activity, we proceed
to check the ability of the network to produce a temporal codification
of the information contained in the stimulus, and 
the robustness of the network response to the
introduction of noise in the input.

In temporal coding, information 
is represented by the timing of action potentials with
respect to an ongoing collective oscillatory pattern of activity. For
instance, when an odour is presented, every neuron in the AL responds
to the odour with some particular timing with  respect to the LFP
\cite{a2}. As a measure of this temporal coding,
we have divided time in periods of the global
average activity, and calculated for each period the
quantity:

\begin{equation}
A_{i}(n) = \frac{1}{C} \int_{T}
[a_{i}(t)-\overline{V(t)}]^{2} dt
\label{aclus}
\end{equation}

\noindent where $i$ represents a particular cluster, 
$n$ a particular period of the mean activity
$\overline{V(t)}$ of the whole network, $a_{i}(t)$ is
the mean activity of cluster $i$, and $C$ is 
an appropiate normalization constant used to get the final 
value of $A_{i}(n)$ in the range $0-1$.

We would expect the coding to be different for each cluster.
In Fig.\ \ref{fig4} we show the results for three different clusters
chosen at random in a network within the SW connectivity regime. It 
can be observed that the activities of the different clusters are 
out of phase and reach their maximum values at different periods of
the global average activity.  

A system with such a coding can rapidly and efficiently perform
computations that are essential to
pattern recognition, and that are much more difficult to perform in a
rate-coding framework \cite{Hopf,Deco,Gabb,Brunel}.
Following this coding scheme, information about an odour is contained
not only in the neural
assembly active at each oscillation cycle, but also in the precise
temporal sequence in which these assemblies are updated during the
response to the odour. Temporal coding thus allows combinatorial
representations in time as well as in space \cite{a2}.

Lastly, in order to check the robustness of the network response,
we have computed correlations between the activities of a given
cluster in five different realizations of the simulation. 
We have introduced a gaussian uncorrelated
noise to the external input: $I^{e}=I_{0}+\sigma \epsilon$, 
where $\sigma$ is the outcome of a normal distibution and
$\epsilon$ is the noise level. The correlations are calculated
as follows:

\begin{equation}
C_{i}^{(rs)}(\epsilon) = \frac{\sum_{n}m_{i}^{(r)}(n)
\cdot m_{i}^{(s)}(n)}
{\sqrt{\sum_{n}m_{i}^{(r)}(n)^{2}
\sum_{n}m_{i}^{(s)}(n)^{2}}}
\end{equation}

\noindent where $r$ and $s$ correspond to different realizations of 
the simulation, $i$ is the cluster, the sums are over all periods $n$,
and the dependence on the noise is implicit in 
the right side of the equation. We have defined the quantity
$m_{i}^{(r)}(n) = A_{i}^{(r)}(n)-\langle A_{i}^{(r)}\rangle$,
with $A_{i}^{(r)}(n)$ the
magnitude from Eq.\ (\ref{aclus}), for a particular realization $r$.
The angle brackets denote average over all periods $n$.

An average of the correlation curves for different pairs of
realizations is plotted in Fig.\ \ref{fig5}.
There are two clearly different regions in the graphic: for small
amounts of noise the correlation is almost $1.0$; 
whereas for a higher noise level the correlation
jumps to a lower value. The discontinuity appears at
a level of noise of approximately $0.1$ percent.

In conclusion, a variety of possible network
topologies has been investigated. Each one gives rise to
different dynamical properties. Regular networks produce
coherent oscillations and temporal coding in a slow time
scale; whereas random networks give rise to fast response
but without coherent oscillations.
We have introduced new results on small-world networks,
showing that both coherent oscillations and temporal coding
can coexist in synergy in a fast time scale. 
At the onset of this research project, intrigued by the olfactory
AL dynamics,  we were searching for a
dynamical system with precisely the properties just described
for the SW networks. Hence the work here reported is not only
interesting from the dynamical systems point of view, but it is
also relevant for the understanding of biological systems. 

We want to acknowledge Gilles Laurent, Alex B\"{a}cker, Maxim 
Bazhenov, Misha Rabinovich and Henry Abarbanel for very insightful 
discusions. 
We thank the Direcci\'{o}n General de
Ense\~{n}anza Superior e Investigaci\'{o}n Cient\'{\i}fica for
financial support (PB97-1448).      
We thank the Universidad Aut\'{o}noma de Madrid for a graduate
fellowship to L. F. L.
and the Centro de Computaci\'{o}n Cient\'{\i}fica (UAM) for 
the computation time.

\begin{figure}
\caption{(a) Characteristic path length $L(p)$ and clustering
coefficient $C(p)$ for the family of randomly rewired graphs,
normalized to the values 
$L(0)$ and $C(0)$ of the regular case.
(b) Average activity oscillation amplitude  $\sigma(p)$ for the
whole range of networks, calculated between $T_{1} = 100$ and 
$T_{2} = 200$. Both curves are averages over ten
realizations of the simulation with parameters $N=797$,
$k=30$ and $g=0.015$. An input signal $I_{0}=1.5$ was injected,
at $t=50$, to $80$ neurons randomly chosen .}
\label{fig1}
\end{figure}

\begin{figure}
\caption{Average activity in a network of $797$ neurons.
(a) Regular network ($p=0.000$).
(b) Small-world network ($p=0.032$).
(c) Random network ($p=1.000$). The input onset occurs at
$t=50$ and is offset at $t=350$. All the parameters
are as described in Fig.\ \ref{fig1}.}
\label{fig2}
\end{figure}

\begin{figure}
\caption{Phase diagram which shows the regions of oscillatory
(clear, high $\sigma$) and nonoscillatory (dark, low $\sigma$) 
activity of the network in the
$(k,p)$ plane. The island that appears on the right side
indicates that the SW (for some range of values of $k$)
is the only regime capable to produce fast
coherent oscillations in the average activity after the 
presentation of the stimulus. 
All parameters are as described in the previous figures.}
\label{fig3}
\end{figure}

\begin{figure}
\caption{(a)-(c) Average activity of three different clusters
of neurons promediated over periods of the global mean activity.
The simulation corresponds with that of Fig.\ \ref{fig2}b,
which lies within the SW region.
(d) Average activity of the whole network showing the coherent 
oscillations over which the activities of clusters are 
promediated.}
\label{fig4}
\end{figure}

\begin{figure}
\caption{Correlation versus noise for two simulations with SW
connectivity. (a) $p=0.032$. (b) $p=0.100$. The rest of parameters
are as described in Fig.\ \ref{fig1}. The plots were obtained as 
follows:
we made five realizations of the simulation for a given
$\epsilon$, and calculated the
point as a double average; first an average over all possible pairs 
of  realizations for a given cluster, and then an average over 
clusters.}
\label{fig5}
\end{figure}

\widetext

\begin{references}

\bibitem{a1}D. J. Watts and S. H. Strogatz, Nature {\bf 393}, 440 
(1998).

\bibitem{barrat}A. Barrat and M. Weigt, submitted to Eur. Phys.
J. B. Also {\em cond-mat/9903411}. 

\bibitem{sw1}M. Barth\'{e}l\'{e}my and L. A. N. Amaral,
Phys. Rev. Lett. {\bf 82}, 3180 (1999).

\bibitem{sw2}M. A. de Menezes, C. F. Moukarzel and T. J. P. Penna, 
submitted to Phys. Rev. Lett. Also {\em cond-mat/9903426}.

\bibitem{sw3}M. E. J. Newman and D. J. Watts, submitted to
Phys. Rev. Lett. Also {\em cond-mat/9903357}.

\bibitem{a2}M. Wehr and G. Laurent, Nature {\bf 384}, 162 (1996).

\bibitem{a3}G. Laurent and H. Davidowitz, Science {\bf 265}, 1872 
(1994).

\bibitem{a5}K. MacLeod and G. Laurent, Science {\bf 265}, 976 (1996). 

\bibitem{a6}K. MacLeod, A. B\"{a}cker and G. Laurent, Nature 
{\bf 395}, 693 (1998).

\bibitem{fer}C. M. Gray, J. Comput. Neurosci. {\bf 1}, 11 (1994).

\bibitem{i1}K. Kaneko, Physica {\bf 23D}, 436 (1986).

\bibitem{i2}D. A. Egolf and H. S. Greenside, Nature {\bf 369}, 851 
(1994). 

\bibitem{i4}I. S. Aranson, D. Golomb and H. Sompolinsky,
Phys. Rev. Lett. {\bf 68}, 3495 (1992).

\bibitem{i6}H. Chat\'{e}, A. Lemaitre, P. Marq and P. Manneville,
Physica {\bf 224A}, 447 (1996). 

\bibitem{i7}A. V. Gaponov-Grekhov and M. I. Rabinovich,
Chaos {\bf 6}, 259 (1996). 

\bibitem{i10}D. Hansel and H. Sompolinsky, Phys. Rev. Lett. {\bf 68}, 
718 (1992).

\bibitem{i11}R. Huerta, M. Bazhenov and  M. I. Rabinovich,
Europhysics Letters {\bf 43(6)}, 719 (1998).

\bibitem{d1}C. van Vreeswijk and H. Sompolinsky,
Science {\bf 274}, 1724 (1996).

\bibitem{Fohlmeister}C. Fohlmeister, W. Gerstner, R. Ritz and J. L. 
van Hemmen,
Neural Comput. {\bf 7}, 1046 (1995).

\bibitem{hhux}A. L. Hodgkin and A. F. Huxley, J. Physiol. {\bf 117},
500 (1952).

\bibitem{11} T. E. Hull, W. H. Enright, B. F. Fellen and R. E. 
Sedgwick,
SIAM J. Num. Anal {\bf 9}, 603 (1972).

\bibitem{Destx}A. Destexhe, Z. F. Mainen and T.J.Sejnowski,
Neural Comput. {\bf 6}, 14 (1993).

\bibitem{Hopf}J. J. Hopfield, Nature {\bf 376}, 33 (1995).

\bibitem{Deco}G. Deco and B. Schurmann, Phys. Rev. Lett. {\bf 79},
4697 (1997).

\bibitem{Gabb}F. Gabbiani, W. Metzner, R. Wessel and C. Koch, Nature
{\bf 384}, 564 (1996).

\bibitem{Brunel}N. Brunel, Network: Computation in Neural Systems {\bf
5}, 449 (1994).

\end{references}
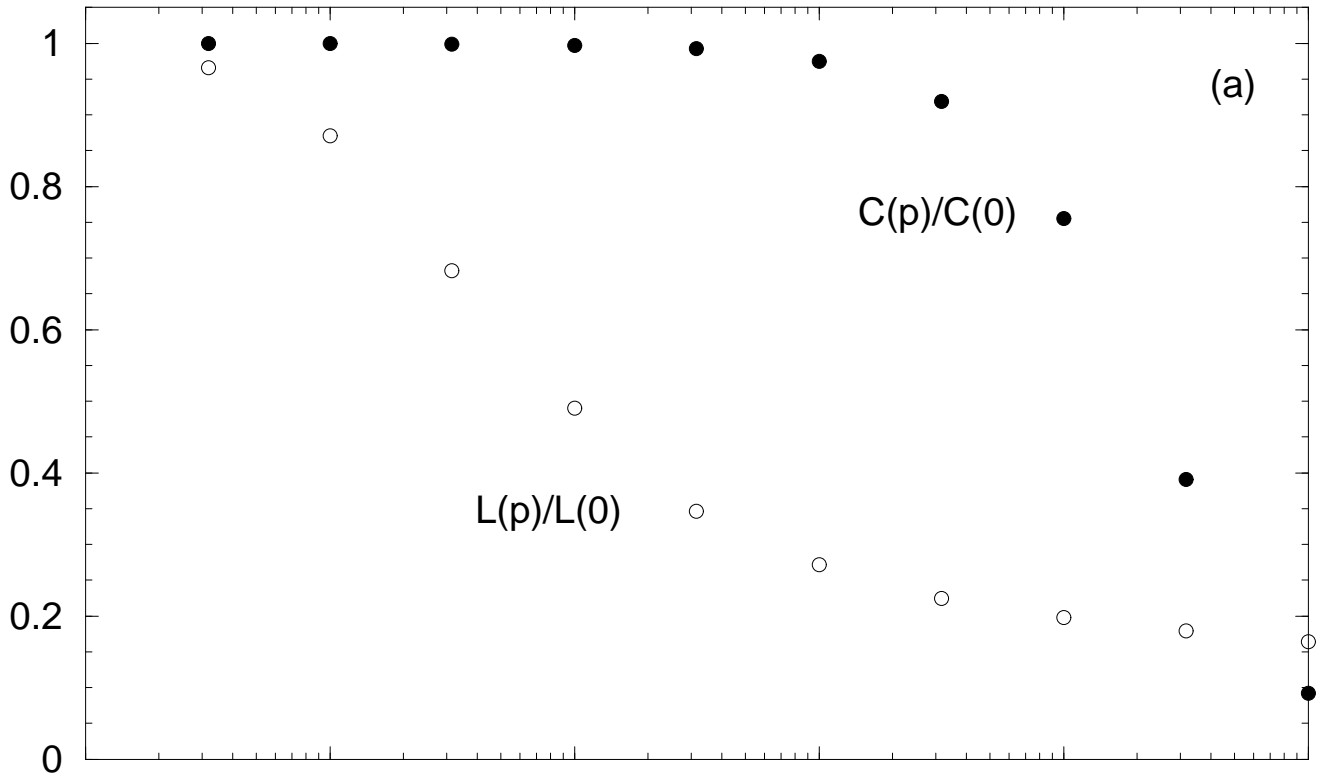
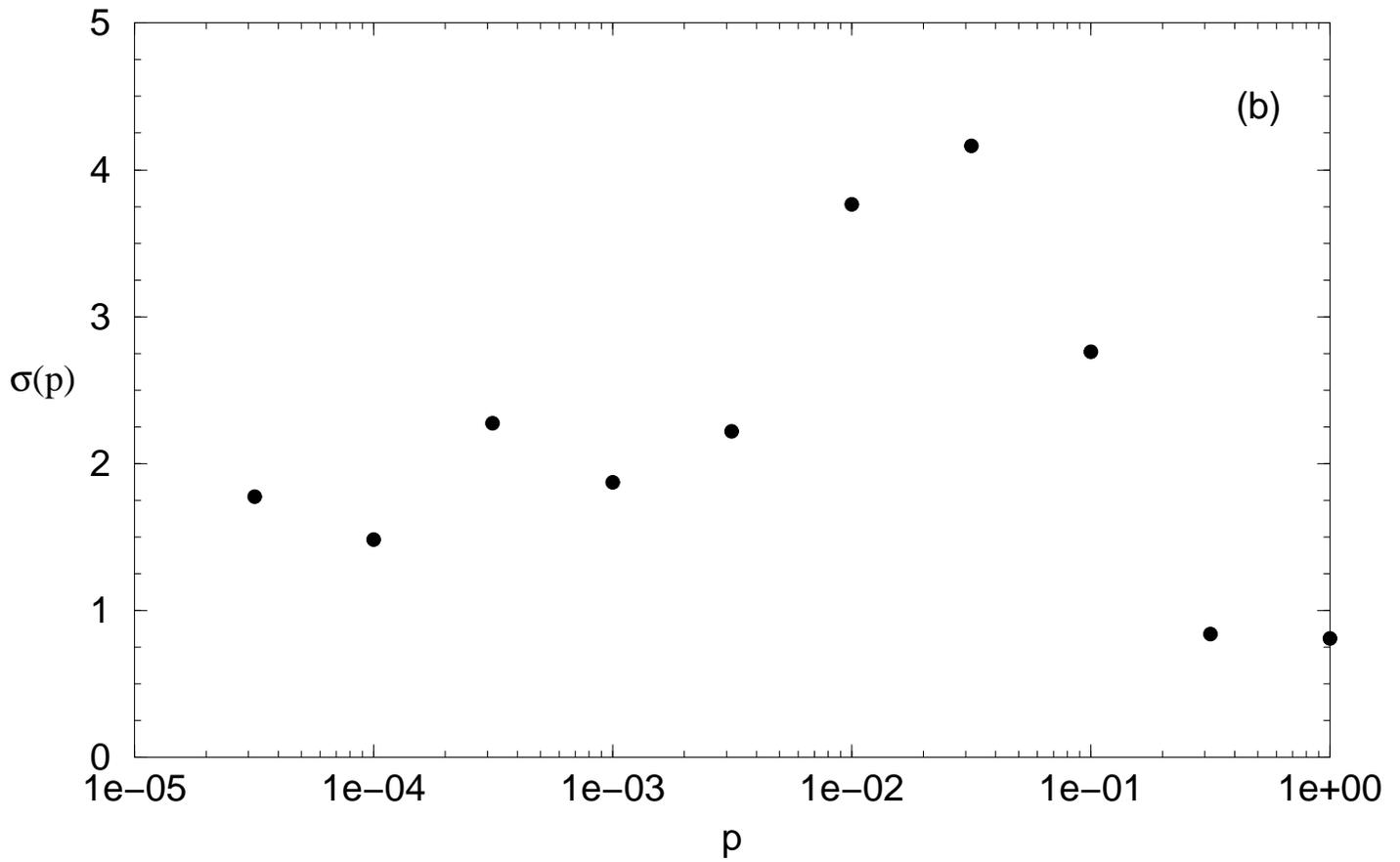
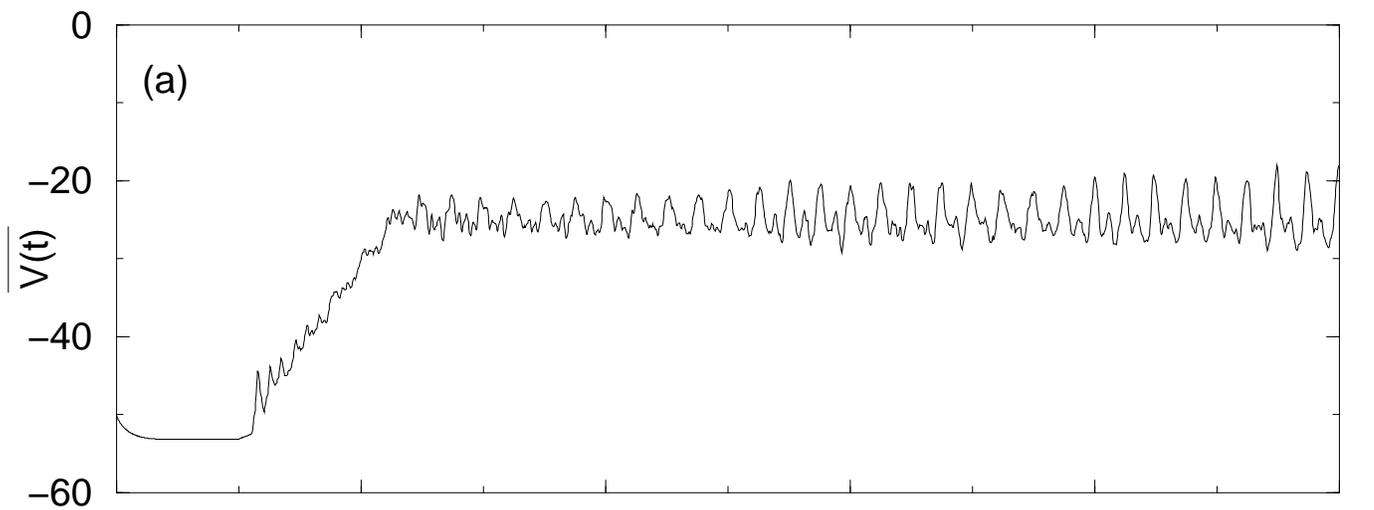
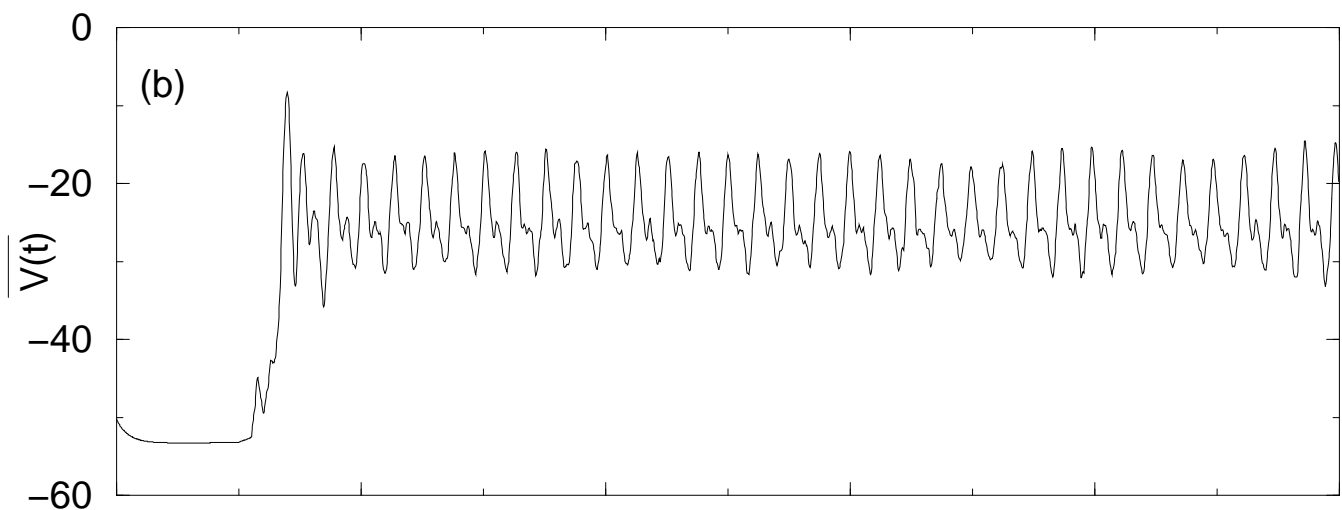
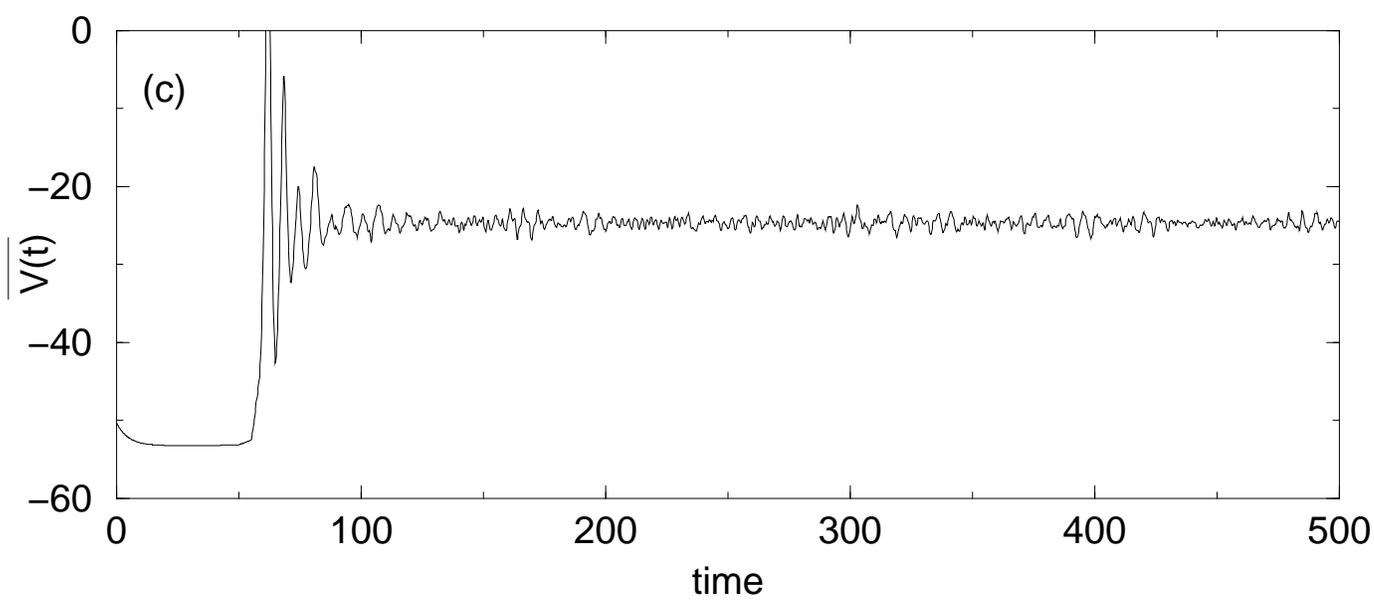
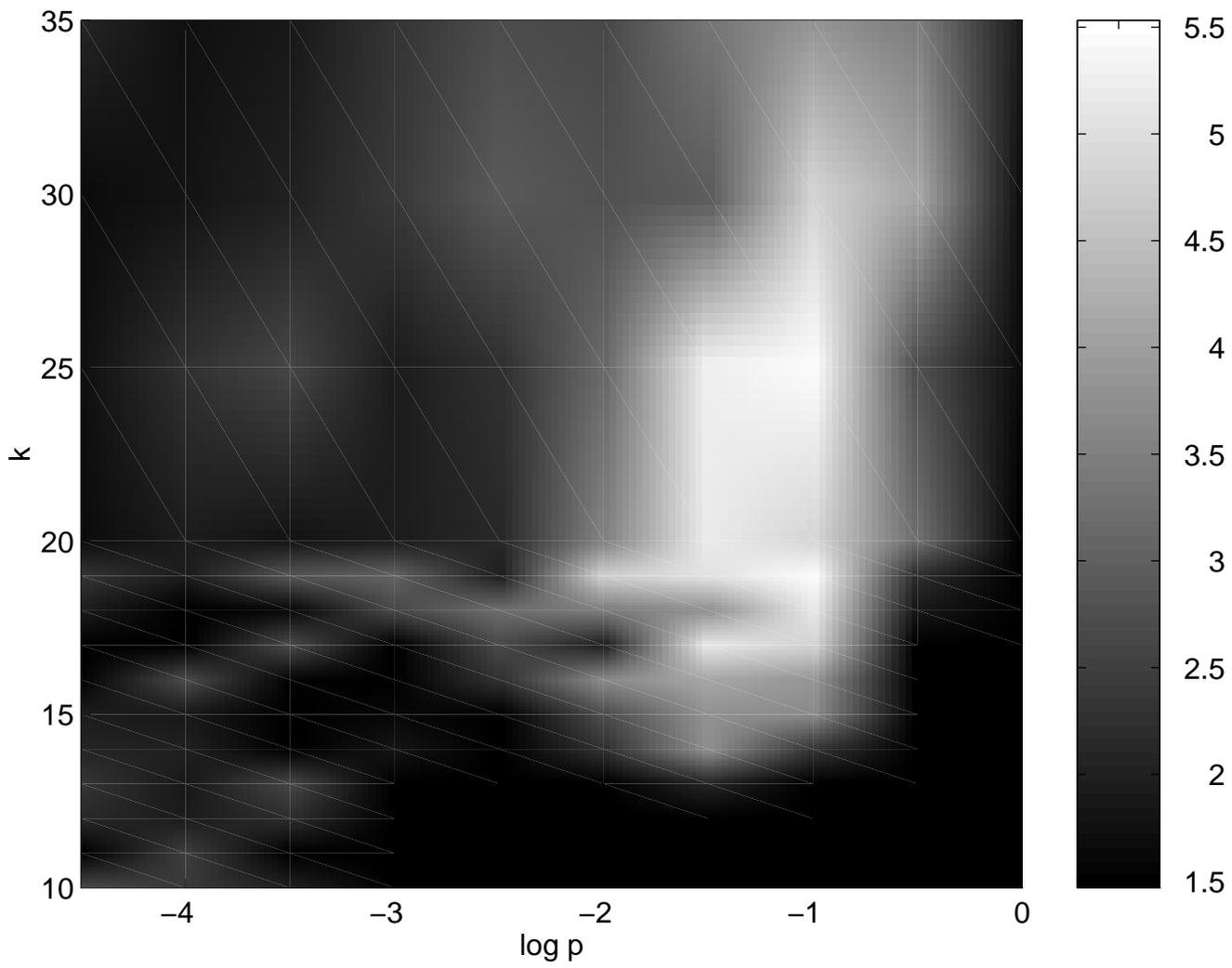
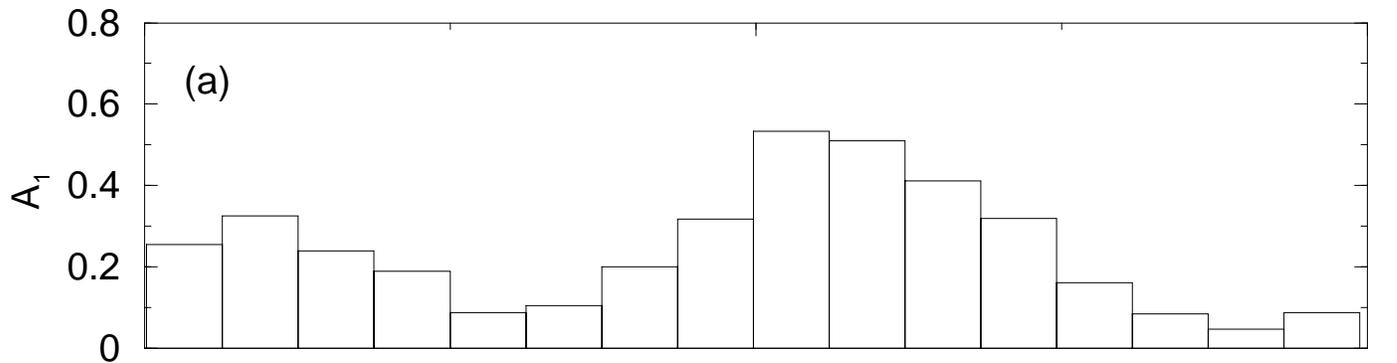
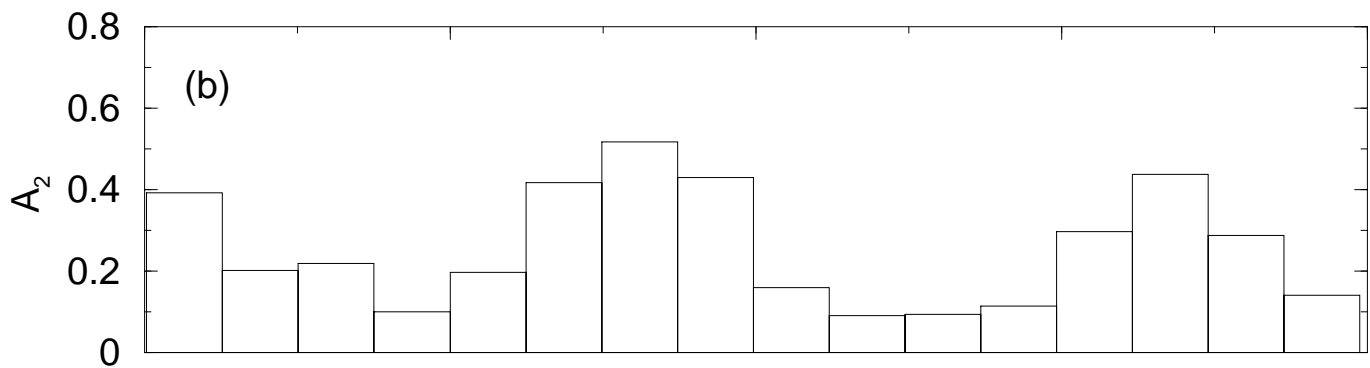
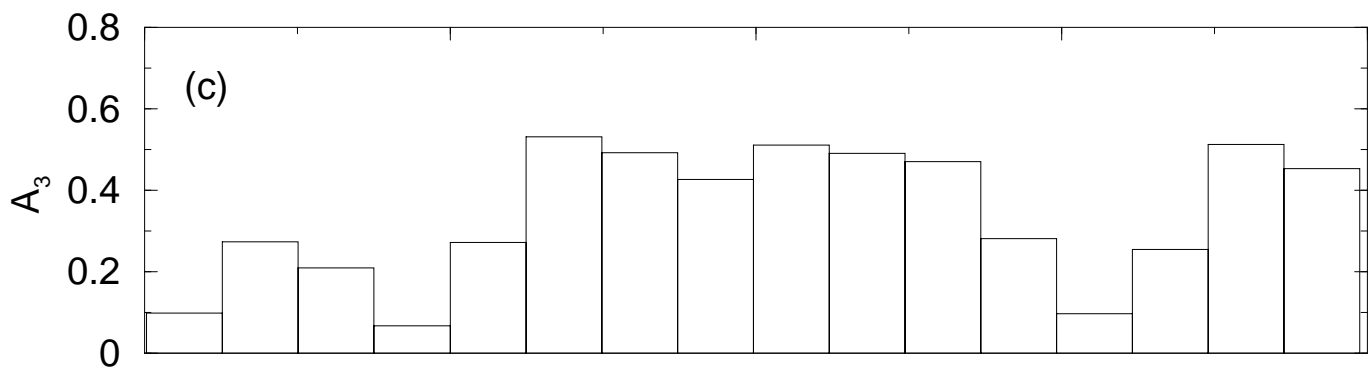
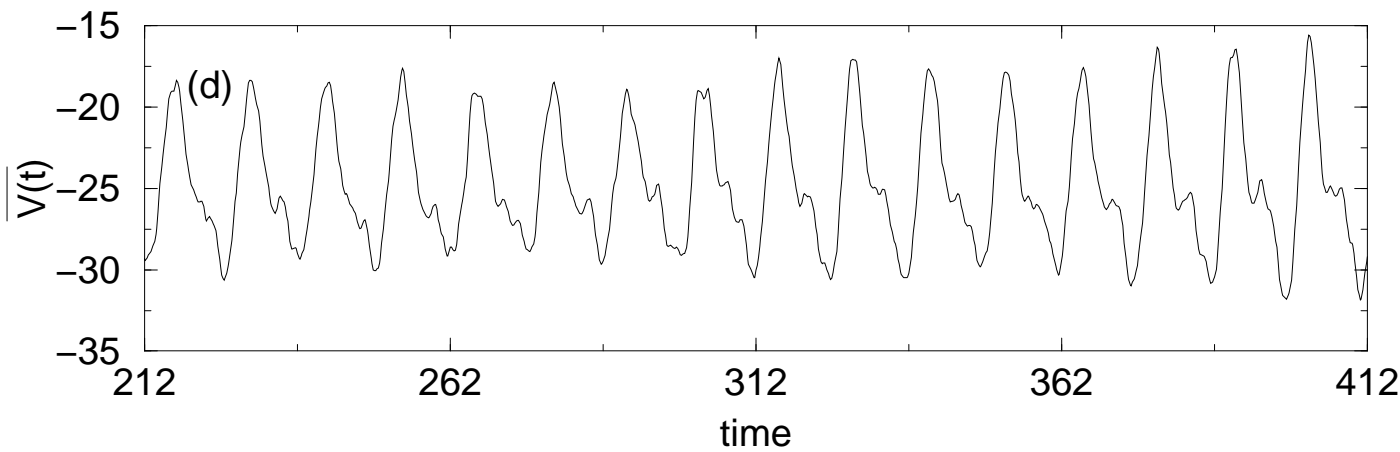
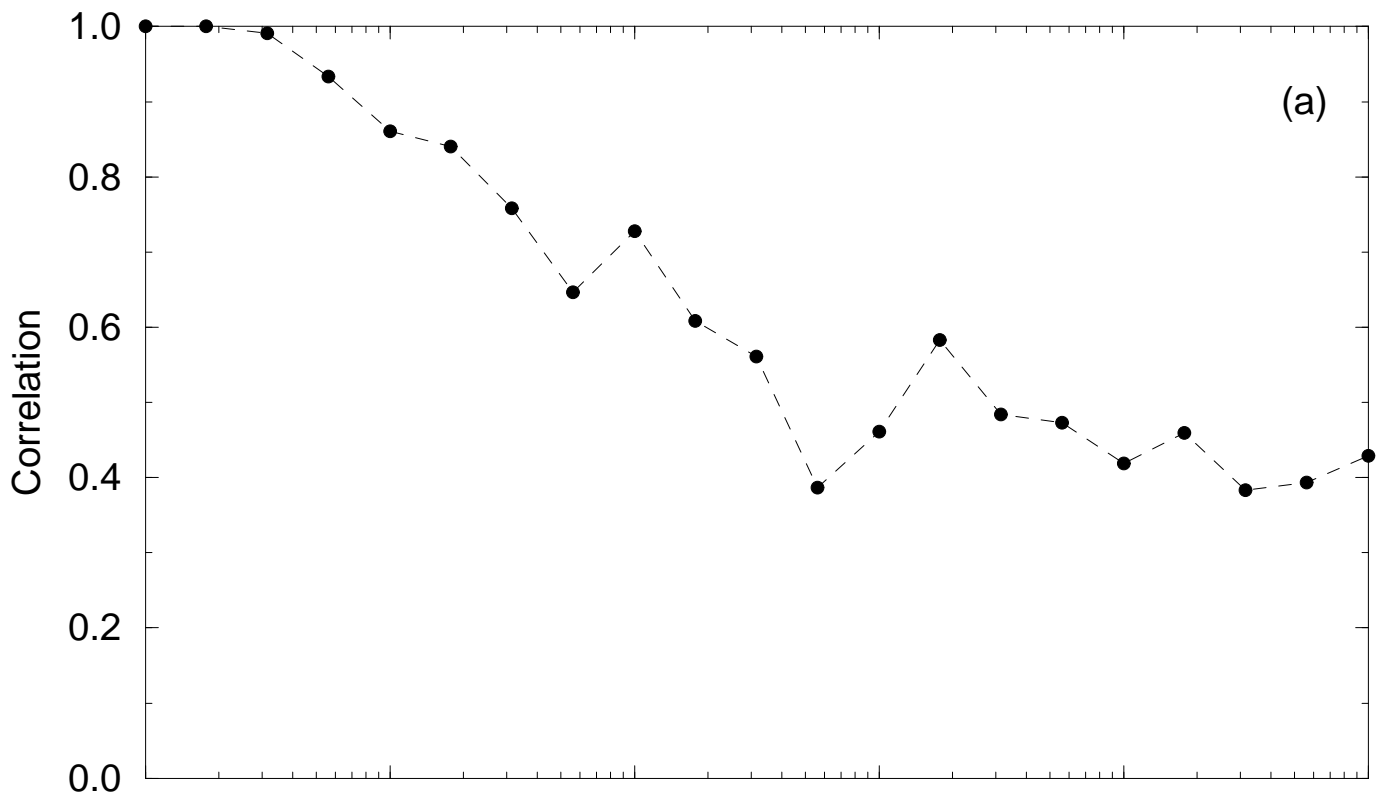
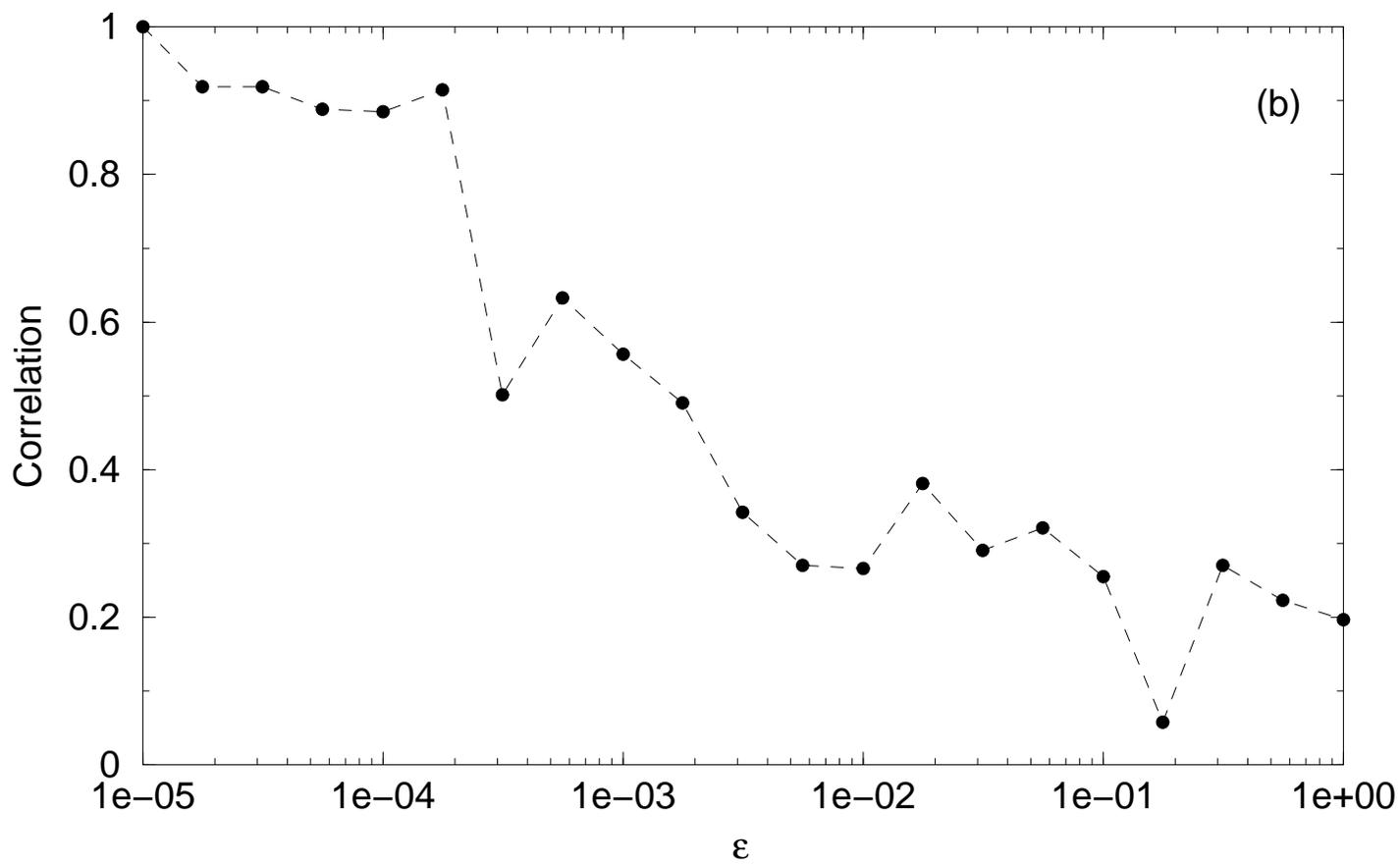
\end{document}